\documentstyle[pre,aps,multicol,amsfonts]{revtex}

\def\R{\relax{\rm I\kern-.18em R}}
\def\Z{\relax{\rm Z\kern-.4em Z}}

\title{ {On the supersymmetric formulation of Unitary Matrix Model of type IIB
}
}
\author{%
Tsukasa Tada$^{a}$ and Asato Tsuchiya$^{b}$}
\address{\  \\
${}^a$ KEK Theory Group \\
1-1 Oho, Tsukuba \\
Ibaraki 305-0801, Japan \\
tada@ccthmail.kek.jp \\
\  \\
${}^b$ Department of Physics, Graduate School of Science \\
Osaka University \\
Toyonaka, Osaka 560-0043, Japan \\
tsuchiya@funpth.phys.sci.osaka-u.ac.jp
}

\date{February, 1999
}

\begin{document}
\draft
\maketitle         
\begin{abstract}
We propose a unitary matrix formulation of type IIB matrix model.
One-loop effective action of the model exhibits supersymmetry for
BPS background in the large N limit.
\end{abstract}

\begin{multicols}{2}
One of the most promising approaches towards a non-perturbative formulation of
string theory is the matrix model
\cite{IKKT,FKKT,BFSS,periwal,yoneya,DVV,itoyama}.
For example, the action of the IIB matrix model \cite{IKKT,FKKT} is nothing but
a reduction \cite{RM} of $ N=1$ ten-dimensional Super Yang Mills theory:
\begin{equation}
S_{\tiny \hbox{IKKT}}=-\frac14\hbox{Tr}(
{[A_\mu,A_\nu]^2})-\frac12\hbox{Tr}{{\bar \psi} {\Gamma^\mu}[A_\mu, \psi] }.
\label{hermite}
\end{equation}
While those matrix models, either so-called Matrix theory in IIA language
or the IIB matrix model (also known as the IKKT model) ,
are composed of
hermite matrices, there are  several regions where models in terms
of unitary matrices would be preferable.

For example, to treat one of the space-time dimensions with periodic
boundary condition, one would have to introduce infinite copies
of the original matrices and end up with introducing a corresponding
gauge field\cite{wati}.  Studies along this line have been
revealing many interesting features of matrix models \cite{wati2}, especially
the relation with  D-branes \cite{KT}. In those studies,
the compactification procedure  is needed to utilize the T-dual
argument.
\footnote{In \cite{AIKKT}, the compactification of matrix models is
considered from a different view of point, that is, rather to explain the
real dimensionality of the world. }
It would be useful if one could have a matrix model in terms of unitary
matrices,
\begin{equation}
U_\mu=e^{i A_\mu} , \label{expa}
\end{equation}
in the first place, since $U_\mu$ apparently represents compactified space (or time). We, then, acquire a coupling constant in the theory, just as
the emergence of the gauge coupling constant in the hermite model upon
conventional compactification.
Therefore, for example, 
it would be interesting to compare the result from the hermite case
 \cite{AIKKT,HNT,KS} with the unitary case in the numerical study.

There is another advantage of using a unitary matrix. In matrix models, pairs
of  matrices which satisfy Heisenberg algebra,
\begin{equation}
[p,q]=\frac{2\pi i}{N} , \label{Heisenberg}
\end{equation}
are used to construct BPS states.
There is no way to represent the above algebra in terms of finite
$N\times N$ hermite matrices. On the other hand,  in terms of
unitary matrices,

\begin{equation}
U=e^{ip} ,  \quad V=e^{iq} ,
\end{equation}
there is a well known representation in terms of finite unitary matrices,
namely the shift and clock matrices:
\begin{equation}
U=\left (\matrix{
0 & 1 & 0 & 0 & 0\cr
0 & 0 & 1 & 0 & 0\cr
0 & 0 & 0 & 1 & 0\cr
\vdots & \vdots & \vdots & \vdots & \ddots\cr
1 & 0 & 0 & 0 & \cdots\cr
}\right ) ,\label{shift}
\end{equation}
\begin{equation}
V=\left (\matrix{
1 & 0 & 0 & 0 \cr
0 & e^{2\pi i/N} & 0 & 0 \cr
0 & 0 & e^{4\pi i/N} & 0 \cr
0 & 0 & 0 & \ddots\cr
}\right ). \label{clock}
\end{equation}
Therefore, we would have an explicit representation of BPS states
in unitary matrix models regularized with finite $N$.
Also, one can speculate that the relation between matrix models and
 noncommutative geometry \cite{Connes} can be more transparent
in this manner.


Despite all of  those advantages that we can anticipate,  there is an
obstruction in building unitary analogue of Matrix theory or the IIB matrix
model.

A naive attempt to rewrite Matrix theory or the IIB matrix model
in unitary matrices,
\begin{equation}
S=-\sum{ \hbox{Tr} ({{U}_{\mu}{U}_{\nu}
{U}^{\dagger}_{\mu}{U}^{\dagger}_{\nu}})} +\frac{i}{2} \hbox{Tr}{{\bar \psi}
{\Gamma^\mu} (U_\mu \psi U_\mu^\dagger
-U_\mu^\dagger \psi U_\mu)},
\label{naive}
\end{equation}
immediately encounters a difficulty  \cite{poly,nishimura}.
In \cite{poly} the unitary version of Matrix theory is considered,  and
it turnes out
that the model does not  possess supersymmetry. This  situation
is the same for the IIB case. The action of the IKKT model (\ref{hermite}) 
has enhanced N=2 supersymmetry:
\begin{equation}
\cases{
\delta^{(1)}\psi =\frac{i}{2} [A_\mu,A_\nu]\Gamma^{\mu\nu}\epsilon \cr
\delta^{(1)}A_\mu =i{\bar \epsilon}\Gamma_\mu\psi ,
}
\end{equation}
\begin{equation}
\cases{
\delta^{(2)}\psi=\xi \cr
\delta^{(2)}A_\mu=0 .}
\end{equation}
However, a  naive generalization of the above algebra to the unitary case,
\begin{equation}
\cases{
\delta^{(1)}\psi =\frac{i}{2}({U}_{\mu}{U}_{\nu}
{U}^{\dagger}_{\mu}{U}^{\dagger}_{\nu}-h.c.) \Gamma^{\mu\nu}\epsilon \cr
\delta^{(1)} U_\mu =i{\bar \epsilon}\Gamma_\mu (U_\mu \psi+\psi U_\mu)
} ,
\cases{
\delta^{(2)}\psi=\xi \cr
\delta^{(2)} U_\mu=0 ,
}  \label{susy}
\end{equation}
does not form the susy algebra.
Since this N=2 supersymmetry in 10 dimensions is
supposed to play an essential role to relate those models with gravity,
the naive unitary models are apparently unsatisfactory.

As a matter of fact, the lack of supersymmetry in the above naive
unitary models is somewhat obvious from the beginning.
When one regards the IKKT model as  the reduction of N=1 
10d Super Yang Mills theory
in the manner of Eguchi-Kawai \cite{RM} to  zero dimension,
they can be seen as a model on a  lattice 
with just one site. The action written
in
terms of unitary matrices is simply the usual plaquett action. Now, fermionic
degrees of freedom doubles in number when fermions are put on a lattice.
Those so-called doublers break the equality in the number of degrees of
freedom between bosons and fermions; thus there is no supersymmetry.

In \cite{nishimura}, it was attempted to resolve this difficulty by
employing the overlap formalism, which has been used in a higher dimensional
lattice model. The overlap formalism enables one to eliminate the unwanted
doubler degree of freedom of the fermions. While this procedure
cuts the degree of freedom of fermions by half, the supersymmetry
is still not manifest in this formalism.

In this note we propose another formulation of a unitary matrix model
of type IIB. To accommodate supersymmetry we simply double
the degrees of freedom of the bosons instead of halving those of the fermions.
\footnote{We owe this idea to H. Kawai.}
Then, the one-loop effective action vanishes for
BPS background in our unitary model.

First, let us calculate the one-loop effective action for the naive unitary
matrix model  (\ref{naive}) for BPS background. We will see  that the
result, itself,  suggests the modification which we propose here. The following
calculation  is almost simple generalization of that of \cite{IKKT} to a
unitary case.

The general classical configuration $U_\mu^0$ for unitary matrices  is
characterized by the existence of a certain set of unitary matrices
$e^{if_{\mu\nu}^0}$, which satisfy the following relations with  $U_\mu^0$:
\begin{equation}
U_\mu^0 U_\nu^0 =U_\nu^0 U_\mu^0 e^{if_{\mu\nu}^0},
\end{equation}
\begin{equation}
[U_\mu^0, e^{if_{\mu\nu}^0}]=0.
\end{equation}
The above relations correspond to the following equations of motion of hermite
matrix model:
\begin{equation}
[A_\mu^0, A_\nu^0]=-i f_{\mu\nu}^0,
\end{equation}
\begin{equation}
[f_{\mu\nu}, A^0_\mu]=0.
\end{equation}

We now put
\begin{equation}
U_\mu=U_\mu^0 e^{iX_\mu}
\end{equation}
and integrate over $ X_\mu$ up to quadratic order.
Expanding the bosonic part of the action,
\begin{equation}
U_\mu U_\nu U_\mu^\dagger U_\nu^\dagger,
\end{equation}
in terms of $ X_\mu$ up to quadratic order yields
\begin{eqnarray}
&&\sum_{\mu \neq \nu} {tr e^{if_{\mu\nu}^0} }  \nonumber\\
&&+ \sum_{\mu, \nu} { tr(X_\mu U_\nu^0 X_\mu U_\nu^{0\dagger}
-X_\mu^2 +  U_\mu^0 X_\nu U_\mu^{0\dagger} X_\nu -X_\nu^2 )e^{i f_{\mu\nu}^0}}
\label{second}\\
&&+tr \left( \sum_\mu \left(U_\mu^0 X_\mu U_\mu^{0\dagger} -X_\mu
\right) \right)^2. \nonumber
\end{eqnarray}
While the first term in (\ref{second}) simply gives  an irrelevant constant
after $X_\mu$ integration, the third term in (\ref{second}) suggests the
following
natural gauge fixing term:
\begin{equation}
F=\sum_\mu (U_\mu^0 X_\mu U_\mu^{0\dagger}-X_\mu).
\end{equation}
The BRS invariant action can be obtained by adding a gauge fixing term and a
ghost term, as follows:
\begin{eqnarray}
S=&-\sum{ \hbox{Tr} ({{U}_{\mu}{U}_{\nu}
{U}^{\dagger}_{\mu}{U}^{\dagger}_{\nu}})} + \hbox{Tr}F^2\nonumber \\
&+\frac{i}{2} \hbox{Tr}{{\bar \psi} {\Gamma^\mu} (U_\mu \psi U_\mu^\dagger
-U_\mu^\dagger \psi U_\mu)}+b\delta_{BRS}F.
\label{BRS}
\end{eqnarray}
The BRS transformation for unitary matrices is
\begin{equation}
\delta_{BRS}U_\mu=e^{-ic}U_\mu e^{ic} -U_\mu=-i[c, U_\mu],
\end{equation}
and for $X_\mu$ it is
\begin{equation}
\delta_{BRS}X_\mu=U_\mu^\dagger c U_\mu+c.
\end{equation}

In the following we  evaluate each quadratic integration for the BPS
background, that is, for the case that $ f^0_{\mu\nu}$ is a {\it c-number}.
It is useful to introduce a basis of $ N\times N$ hermite matrices $T^a$  which
satisfies $tr T^a T^b = \delta_{ab}$. In terms of this basis $X_\mu$ can be
written as
\begin{equation}
 X_\mu =\sum T^a x^a_\mu.
\end{equation}
Also, we define the following matrix:
\begin{equation}
(u_\mu^0)_{{ab}}\equiv \hbox{Tr}T^{ a} U_\mu^0 T^{ b} U_\mu^{0\dagger}.
\end{equation}
It is easy to see that $ u_\mu^0$ is a $ N^2\times N^2$ unitary matrix.
Note that $u_\mu^0$ commutes each other,
\begin{equation}
u^0_\mu u^0_\nu =u^0_\nu u^0_\mu ,\label{commute}
\end{equation}
if $ U_\mu^0$ corresponds to the BPS background.

First, (\ref{second}), the quadratic part of $-\sum{ \hbox{Tr}
({{U}_{\mu}{U}_{\nu} {U}^{\dagger}_{\mu}{U}^{\dagger}_{\nu}})} + \hbox{Tr}F^2$
becomes
\begin{equation}
\sum_{\mu, \nu} { tr(X_\mu U_\nu^0 X_\mu U_\nu^{0\dagger}
-X_\mu^2 +  U_\mu^0 X_\nu U_\mu^{0\dagger} X_\nu -X_\nu^2 )\cos{f_{\mu\nu}^0}},
\end{equation}
provided that $ f^0_{\mu\nu}$ is a {\it c-number}.
This can be written in terms of the notation introduced above,
\begin{equation}
-\sum_{\mu, \nu}\cos{f_{\mu\nu}^0}\left({}^t x_\mu u_\nu^0 x_\mu +{}^t x_\mu
u_\nu^{0\dagger}  x_\mu -2 {}^t  x_\mu x_\mu  \right).
\end{equation}
The integration over $x_\mu$ is evaluated as
\begin{equation}
\prod_{\mu}{ \sqrt{\det \sum_\nu \left[
\left( 2-u_\nu^0 -u_\nu^{0\dagger} \right) \cos{f^0_{\mu\nu}}\right] }}.
\end{equation}
Secondly, the quadratic part of the ghost term,
\begin{equation}
b\sum_\mu ( U_\mu^0 c U_\mu ^{0\dagger} +U_\mu ^{0\dagger} c U_\mu ^{0}
-2c ),
\end{equation}
can be evaluated in a  similar manner,  and leads to
\begin{equation}
\det \sum_\nu \left( 2-u_\nu^0-u_\nu^{0\dagger} \right) .
\end{equation}
Finally, expanding the fermion matrices with the same basis,
\begin{equation}
\psi=\sum T^a \psi_a,
\end{equation}
gives the following expression for the fermionic term in the action:
\begin{equation}
\frac{i}{2} \hbox{Tr}{{\bar \psi} {\Gamma^\mu} (U_\mu \psi U_\mu^\dagger
-U_\mu^\dagger \psi U_\mu)}=\frac{i}{2}\psi_a ^\dagger\sum_\mu \Gamma^0
(\Gamma^\mu u_\mu-\Gamma^\mu u_\mu^\dagger)\psi_a .
\end{equation}
The integration over ten-dimensional Majorana-Weyl fermions yields
\begin{equation}
\det\left(\frac{i}{2} \sum_\mu \Gamma^0\Gamma^\mu \left(u_\mu-u_\mu^\dagger
\right)\right)^{ 8} .
\end{equation}
It is rather convenient to rewrite  the above while noting  that
\begin{equation}
\det\left(\frac{i}{2} \sum_\mu \Gamma^0\Gamma^\mu \left(u_\mu-u_\mu^\dagger
\right)\right)
=\sqrt{\det\left( \frac{i}{2} \sum_\mu\Gamma^0\Gamma^\mu (u_\mu-u_\mu^\dagger
)\right)^2}.
\end{equation}
With a simple gamma matrix algebra and (\ref{commute}) one finds
\begin{equation}
\det\left(\frac{i}{2} \sum_\mu \Gamma^0\Gamma^\mu \left(u_\mu-u_\mu^\dagger
\right)\right) =
\sqrt{\det \left[
-\frac14 \left( u_\lambda^0-u_\lambda^{0\dagger} \right)^2  \right]}.
\end{equation}
Collecting all of the above result,  we obtain the following one-loop partition
function
for general BPS background:
\begin{eqnarray}
&&\prod_{\mu}{ \left(\det \sum_\nu \left[
\left( 2-u_\nu^0 -u_\nu^{0\dagger} \right) \cos{f^0_{\mu\nu}}\right]
\right)^{-1/2}} \nonumber \\
&&\times  \left(\det \sum_\nu \left( 2-u_\nu^0-u_\nu^{0\dagger} \right) \right)
\left(\det \left[
-\frac14 \left( u_\lambda^0-u_\lambda^{0\dagger} \right)^2  \right]
\right)^{4}.
\label{GBPS}\end{eqnarray}

The above quantity is supposed to become a constant if it is the effective
partition function for supersymmetric background. Since we do not have
supersymmetric algebra for this model, it is rather the expected result.
However, the above result has a very suggestive form to obtain a supersymmetric
model.

First, obstruction to supersymmetry in (\ref{GBPS}) is $\cos{f^0_{\mu\nu}} $
term in the bosonic contribution. Without it, the bosonic contribution would
have the same form as the ghost contribution. To avoid that difficulty,
we perform the following analysis in $ 1/N$ expansion.
We postulate that the magnitude of $ f_{\mu\nu}^0$ for the BPS state is
\begin{equation}
f_{\mu\nu}^0 \sim O(\frac1N).  \label{order}
\end{equation}
Eqs. (\ref{shift}) and  (\ref{clock}) provide
 an  example of non-commuting pair
of unitary matrices that satisfy (\ref{order}), which may seem to lead
commuting pair of matrices in the large N limit at first sight.
Then, in the $ 1/N$ expansion
\begin{equation}
\cos f_{\mu\nu}^0 =1+ O(\frac{1}{N^2})
\end{equation}
for the general BPS background. Note that for the background of
diagonal matrices, that is, for the $ f_{\mu\nu}^0=0$ case, $ \cos f_{\mu\nu}^0
=1$
without the subleading contribution of $ 1/N$.
Also, we expect the contribution of $ O(1/N)$ from the integration
over fermion zero modes.
Therefore, the one-loop partition function up to the leading 
order of $ 1/N$ is
 \begin{equation}
 \left\{\det \sum_\nu \left( 2-u_\nu^0-u_\nu^{0\dagger} \right)
\right\}^{({-5+1})} \left(\det \left[
-\frac14 \left( u_\lambda^0-u_\lambda^{0\dagger} \right)^2  \right] 
\right)^{4}. \label{one-loop}
\end{equation}
One can now see that the above expression would become a constant by
cancelling each factor from the denominator and numerator if
$ u_\nu^0$ were replaced by $ (u_\nu^0)^2$.
In fact, this is possible by
replacing $U_\mu$ with $U_\mu^2$ in the bosonic action and gauge
fixing term, as follows:
\begin{equation}
\sum{ \hbox{Tr} ({{U}_{\mu}^2 {U}_{\nu} ^2 ({U}^{\dagger}_{\mu})^2
({U}^{\dagger}_{\nu})^2})} ,
\end{equation}
\begin{equation}
F=\sum_\mu ((U_\mu^0)^2 X_\mu (U_\mu^{0\dagger})^2-X_\mu).
\end{equation}
We then expand $ U_\mu$ as
\begin{equation}
U_\mu U_\mu =U_\mu^0 U_\mu^0 e^{iX_\mu},
\end{equation}
where the summation over $ \mu$ is not taken.
This procedure corresponds to integration with the measure $ {\cal D}(U_\mu )^2
$. Then, the contribution from the bosonic part of the action
and the gauge fixing term yield in the leading $ 1/N$ expansion
\begin{equation}
{}^t x_\mu \left( u_\nu^2  +(u_\nu^\dagger )^2  -2 \right) x_\mu +O(\frac1N),
\end{equation}
while ghost part contributes as
\begin{equation}
 \hbox{Tr}b ( \sum_\mu (U_\mu^0)^2 c (U_\mu^{0\dagger})^2
+(U_\mu^{0\dagger})^2 c (U_\mu^{0})^2 -2c ).
\end{equation}
The above two give the contribution to the partition function,
\begin{equation}
 \left\{\det \sum_\nu \left( 2-\left(u_\nu^0 \right)^2-
\left(u_\nu^{0\dagger}\right)^2 \right) \right\}^{({-4})},
\end{equation}
which cancels  with the fermionic contribution in (\ref{one-loop}) up
to an irrelevant numerical factor of $ \frac14$.
Here, we have used the completeness of the Hermite basis,
$(u^2)_{ac}=u_{ab}u_{bc}=\hbox{Tr}T^a U^2 T^c (U^\dagger)^2 $.
Taking care of the numerical factor $ \frac14$ by normalizing the fermionic
part, we obtain
a unitary matrix model, which exhibits supersymmetry at the one-loop level in
the large N limit:
\begin{eqnarray}
&&\int {\cal D} (U_\mu)^2 {\cal D} \psi \exp [ \sum{ \hbox{Tr} ({{U}_{\mu}^2
{U}_{\nu} ^2 ({U}^{\dagger}_{\mu})^2 ({U}^{\dagger}_{\nu})^2})}
\nonumber \\ && \qquad \qquad \qquad -i \sum
\hbox{Tr}{{\bar \psi} {\Gamma^\mu} (U_\mu \psi U_\mu^\dagger
-U_\mu^\dagger \psi U_\mu)}]. \label{main}
\end{eqnarray}

The above expression (\ref{main}) summarizes the main result of
the present paper. A few remarks are in order:
First, the above expression vanishes for the BPS states only in the 
large N limit. For example, consider the case $ f_{\mu\nu}^0 =0$; namely,
 the case that $ U_\mu^0$ commutes with each other. In that case that 
there exist $N $ fermion zero modes to be integrated out,  which 
 produce non-trivial interactions between the eigenvalues of the matrices
in the hermite case \cite{AIKKT,GG}. Since total number of integration 
is $ N^2$, the contribution from those zero modes yields $ O(1/N)$,
nonvanishing subleading term. 
The analysis of fermion zero modes for more general backgrounds 
involves some technical issues so we left it for future study.
Second, we were unable to find a supersymmetric transformation of the
classical action of (\ref{main}). Therefore, the supersymmetry shown here
seems to be only at the one-loop level.  It would be interesting to see whether supersymmetry persists at the two-loop order. Also, the calculation
for a more general background like the block diagonal background
is rather straightforward.
Third, there will be other mappings from hermite matrices to unitary ones
than that in (\ref{expa}). For example, $(I+iA_\mu /2)(I-iA_\mu /2)^{-1}$ also
leads to a unitary matrix. In this case, however, the relation between
the hermite model and the unitary one becomes more complicated.
Lastly, we seem to need the postulation (\ref{order}) to obtain
the supersymmetric states, that is, BPS state.
Since the postulation (\ref{order}) is related with the scaling which
we are taking, the better understanding of the physical meaning
of the above postulation would be  desired.

In summary we have proposed a new formulation of the IIB matrix model
in terms of unitary matrices, (\ref{main}). The model has a vanishing
 one-loop effective action for the BPS states in the large N limit.
 We hope that we can shed more light on the issues addressed 
above in future publications.

{\bf Acknowledgments}: We would like to thank
H. Kawai  for the collaboration at the early stage
of the project and for illuminating discussions.
We have also greatly benefitted from countless discussions with H. Aoki, N.
Ishibashi, S. Iso, H. Itoyama, Y. Kitazawa, T. Suyama 
and colleagues in the KEK theory group.

\end{multicols}
\end{document}